\documentclass[twocolumn,showpacs,preprintnumbers,amsmath,amssymb]{revtex4}

\usepackage{epsfig}

\newcommand{\be}{\begin{equation}}
\newcommand{\ee}{\end{equation}}
\newcommand{\ba}{\begin{array}}
\newcommand{\ea}{\end{array}}

\newcommand{\tr}{\mbox{Tr}}                                        
\newcommand{\bra}[1]{\ensuremath{\langle #1 |}}
\newcommand{\ket}[1]{\ensuremath{| #1 \rangle}}

\newcommand{\mathbbm}{\bf}

\begin{document}

\preprint{flo/abu/marek}

\title{Concurrence of mixed bipartite quantum states in arbitrary dimensions}

\author{Florian Mintert$^{1,2}$, Marek Ku\'s$^2$, and Andreas Buchleitner$^1$}

\affiliation{
  $^1$Max-Planck-Institut f\"ur Physik komplexer Systeme,
  N\"othnitzerstr. 38, D-01187 Dresden
}

\affiliation{
 $^2$Centrum Fizyki Teoretycznej, Polska Akademia Nauk,
  Al.~Lotnik{\'o}w 32/44, PL-02-668 Warszawa
}

\date{\today}
\begin{abstract}
We derive a lower bound for the concurrence of mixed bipartite quantum states,
valid in arbitrary dimensions. As a corollary, a weaker, purely algebraic
estimate is found, which detects mixed entangled states with positive partial transpose.
\end{abstract}


\maketitle

In classical physics, one can always divide a system into subsystems,
such that complete information on the entity implies a complete 
description of 
its individual parts,
and vice versa.
In quantum physics, 
this no longer holds true: whilst one can still divide a
system into subsystems,
a complete description of the system state in terms of a pure
state does not necessarily assign a pure state to each subsystem.
The subsystems of generic pure states
are correlated in a way without
classical analog -- they are {\em entangled}.

While such quantum entanglement arguably incarnates the key difference between
the quantum and the classical world, and is nowadays understood as a
resource in various tasks of quantum information processing \cite{nielsen00} 
such as
cryptography, teleportation, and quantum computation, it remains hard to
quantify, for arbitrary quantum states \cite{bru02}. In particular, when
coupled to an 
environment, pure quantum states rapidly evolve into mixed states which bear
entanglement {\em together} with classical probabilistic correlations, and the
latter have to be distinguished from the former.
Furthermore, the complete characterization of the nonclassical
correlations of a given state becomes an ever more complex task as the Hilbert
space dimension increases, thus turning into a computationally extremely
intricate problem.  

No equally versatile as computationally manageable
entanglement measure for mixed states  
is available so far, although various more or less
pragmatically motivated quantities have been proposed. The most popular
indicator of entanglement is the positive partial transpose (ppt) criterion
\cite{per96} and variants thereof, such as negativity \cite{vidal02}, though
these do not reliably detect arbitrary entangled states. Another approach for
quantifying entanglement is through entanglement witnesses \cite{ter00b} 
which, however,
need to be constructed anew for each given quantum state, and such
construction can be rather involved \cite{doh02}. Finally, there are mixed
state generalizations of pure state entanglement measures
\cite{wot93,ben96,wot98,uhl00,run01}, 
which, in general, require a high dimensional optimisation procedure.
By construction, any numerical
evaluation of these latter quantities only yields upper bounds for the
entanglement of a given state but {\em cannot} reliably distinguish it from
separable states, let alone provide a reliable {\em quantitative} estimate of
the state's actual degree of entanglement. 

In the present Letter,   
we improve on that situation: We
derive a {\em lower} bound of concurrence \cite{wot93,wot98} -- a quantity 
which is strictly larger than
zero for nonvanishing entanglement -- of mixed bipartite quantum states in
arbitrary dimensions. Our bound is given by a purely algebraic expression
which is readily evaluated for arbitrary states, and can be
tightened numerically on a relatively low-dimensional parameter space, of
reduced dimension as compared to hitherto available optimisation procedures
\cite{uhl98}. This complements already available upper bounds
\cite{uhl00,run01} and provides, for the first time, a rather precise estimate
of the actual value of concurrence.

We start out with the definition \cite{run01} of a pure state's concurrence 
as
$c(\psi)=\sqrt{|\langle\psi|\psi\rangle|^2-\tr\varrho_r^2}$,
where the reduced density matrix $\varrho_r$ is obtained by tracing over
one subsystem. The concurrence of 
mixed states $\varrho$ is then
given as 
the convex roof
\be
c(\varrho)=\inf \sum_i p_i c(\Psi_i), \hspace{.2cm}
\varrho=\sum_i p_i \ket{\Psi_i}\bra{\Psi_i}, \hspace{.2cm}
p_i \ge 0,
\label{concdef}
\ee
of all possible decompositions into pure states 
$\ket{\Psi_i}$. Consequently,
%Therefore, with the above definition,
$c(\varrho)$ vanishes if and only if $\varrho$ exhibits purely classical 
correlations, i.e. if the state is {\em separable} and hence can be
represented as a convex sum over product states,
$\varrho=\sum_ip_i\varrho_i^{(1)}\otimes\varrho_i^{(2)}$, 
with $p_i\ge 0$, and
$\varrho_i^{(1)}$ and $\varrho_i^{(2)}$ states
on the subcomponents ${\cal H}_1$ and ${\cal H}_2$ of the total Hilbert space 
${\cal H}={\cal H}_1\otimes{\cal H}_2$.
Given the dimensions $n_1$ and $n_2$ of ${\cal H}_1$ and ${\cal H}_2$,
respectively, eq.~(1) defines a high dimensional optimisation problem which is
rather cumbersome to solve. Furthermore, as already mentioned above, 
such optimisation can only yield an
upper bound for $c(\rho)$, by virtue of the definition of the infimum. 

To estimate $c(\rho)$ from below, we first replace, for
convenience, the $\ket{\Psi_i}$ by the subnormalized states
$\ket{\psi_i}=\sqrt{p_i}\ket{\Psi_i}$ in eq.~(\ref{concdef}).
Given a valid decomposition $\{
\ket{\phi_i} \}$ of $\varrho$ into subnormalized states, any other suitable
set $\{ \ket{\psi_i} \}$ is obtained \cite{schroed} by transformations
$V\in{\mathbbm C}^{N\times r}$, with $r$ and $N$ the lengths of the sets $\{
\ket{\phi_i} \}$ and $\{ \ket{\psi_i} \}$, respectively, 
\be
\ket{\psi_i}=\sum_{j=1}^r V_{ij} \ket{\phi_j}\ ,\
\sum_{i=1}^NV^\dagger_{ki}V_{ij}=\delta_{j,k}\ . 
\label{allens}
\ee
It is now crucial to realize that the concurrence of a
pure state $\ket{\psi}$ can be expressed as the square root of the function 
\begin{eqnarray}
f(\psi_1,\psi_2,\psi_3,\psi_4)=
\langle\psi_2|\psi_1\rangle\langle\psi_4|\psi_3\rangle- \nonumber\\
\tr_1\left(
\left(\tr_2\ket{\psi_1}\bra{\psi_2}\right)
\left(\tr_2\ket{\psi_3}\bra{\psi_4}\right)\right)\ ,
\label{function}
\end{eqnarray}
evaluated at $\psi=\psi_1=\psi_2=\psi_3=\psi_4$,
where $\tr_1$ and $\tr_2$ denote the traces over the first and the second
subsystem.
$f$ is linear in 
its first and third, and anti-linear
in the second and fourth argument.
Due to these properties the definition (\ref{concdef}) can be
reformulated as an infimum over transformations $V$:
\be
c(\varrho) =\inf_V{\cal C},
\hspace{.3cm}\mbox{with}\hspace{.3cm}
{\cal C}=\sum_{i=1}^N
\left(\left[ V\otimes V A \hspace{.1cm}
V^\dagger\otimes V^\dagger\right]_{ii}^{ii}\right)^\frac12\ .
\label{conc}
\ee
Herein, the tensor $A$, defined by 
$A_{jk}^{lm}=f(\phi_j,\phi_l,\phi_k,\phi_m)$ \cite{bad02} is
hermitian,
$A_{jk}^{lm}=(A_{lm}^{jk})^\ast$, and
symmetric with respect to a simultaneous exchange of
both its co- and contravariant indices $A_{jk}^{lm}=A_{kj}^{ml}$.
Due to the symmetry of the transformation $V\otimes V$ under exchange
of the subsystems of $A$, 
we can replace
$A_{jk}^{lm}$ in eq.~(\ref{conc}) by the symmetrised elements
\be
{\cal A}_{jk}^{lm}=\frac12\left(A_{jk}^{lm}+A_{kj}^{lm}\right) \ ,
\ee
which is equivalent to a symmetrisation over both subsystems in eq. (\ref{function}).
It can be shown that ${\cal A}$ is positive semidefinite and that its
support lies in
an antisymmetric subspace,
{\it i.e.}, all elements of ${\cal A}$ with respect to fully symmetric
linear combinations of product states vanish.
Since the antisymmetric subspace has dimension $m=n_1(n_1-1)n_2(n_2-1)/4$,
${\cal A}$ 
has at most $m$ non-vanishing eigenvalues.

Due to the discussed symmetries ${\cal A}$ can be expanded in a basis of real symmetric matrices
$\Lambda^\alpha\in{\mathbbm R}^{r\times r}$
\be
{\cal A}_{jk}^{lm}=\sum_{\alpha,\beta}B_{\alpha\beta}\hspace{.1cm}
\Lambda_{jk}^\alpha\Lambda_{lm}^\beta ,
\label{expandA}
\ee
with $B$ hermitian and positive semi definite.
With the eigenvalues and associated eigenvectors of $B$
($B\vec x^\alpha=\mu_\alpha\vec x^\alpha$ and
$\vec x^\alpha=[x^\alpha_1,\hdots,x^\alpha_i,\hdots]$)
we can construct a properly normalised eigensystem
$T^\alpha$ of ${\cal A}$
\be
T_\alpha=\sqrt{\mu_\alpha}\sum_\beta x_\beta^\alpha
e^{i\phi_{\alpha}}\Lambda^\beta ={\cal T}_{\alpha}e^{i\phi_{\alpha}} \ ,\
\alpha=1,\ldots,m.
\ee
We explicitly take into account the free phase factors
$\exp(i\phi_{\alpha})$,
as they will be crucial in the following.
Consequently,
\be
{\cal A}_{jk}^{lm}=\sum_\alpha
T_{jk}^\alpha\left(T_{lm}^\alpha\right)^\ast = \sum_\alpha {\cal
T}_{jk}^\alpha\left({\cal T}_{lm}^\alpha\right)^\ast\ .
\ee
Hence, eq.~(\ref{conc}) can now be rewritten as
\be
{\cal C}=\sum_{i=1}^N\Bigl(\sum_\alpha
\Bigl|\Bigl[VT^\alpha V^T\Bigr]_{ii}\Bigr|^2
\Bigr)^\frac12 \ ,
\label{capc}
\ee
the infimum of which gives the concurrence of the mixed state $\varrho$.

Note that eq.~(\ref{capc}) resembles the {\em concurrence vector} introduced
in \cite{aud01}, a quantity with elements analogous to 
$\sum_{i=1}^N\left|\left[VT^\alpha V^T\right]_{ii}\right|$.
In \cite{aud01} it was shown that the concurrence vector vanishes identically,
for suitably chosen $V$, 
if and only if $\varrho$ is separable, thus providing a separability criterion.
Whilst the very same equivalence holds for vanishing ${\cal C}$,
the explicit expression (\ref{capc}) allows us to proceed further:
The Cauchy-Schwarz inequality
$\left(\sum_\alpha x_\alpha^2\right)^\frac12
\left(\sum_\alpha y_\alpha^2\right)^\frac12\ge
\sum_\alpha x_\alpha y_\alpha$, and
$\sum_\alpha|z_\alpha|\ge|\sum_\alpha z_\alpha|$, for
$z_\alpha\in{\mathbbm C}$,
imply
\be
c(\varrho)\ge\inf_V \sum_{i=1}^N
\Bigl|\Bigl[V
\Bigl(\sum_\alpha z_\alpha{\cal T}^\alpha\Bigr)V^T
\Bigl]_{ii}\Bigr| ,
\label{lowb}
\ee
for any set $z_\alpha =y_\alpha \exp(i\phi_{\alpha})$, with $y_\alpha\ge 0$, 
$\sum_\alpha y_\alpha^2=1$.
The infimum on the rhs is given by
$\lambda_1-\sum_{i>1}\lambda_i$,
where $\lambda_j$ are the singular values of
${\cal T}=\sum_\alpha z_\alpha {\cal T}_\alpha$, i.e., the square roots of the
eigenvalues of the 
positive hermitian matrix ${\cal T}{\cal T}^{\dagger}$
in decreasing order \cite{uhl00}.
Hence, we arrive at the desired lower bound, 
\be
c(\varrho)\ge\lambda_1-\sum_{i>1}\lambda_i ,
\label{bound}
\ee
with 
the $\lambda_j$ dependent on the choice of the $y_\alpha$ and
$\phi_\alpha$.

Note that each set 
$\{ y_{\alpha}, \phi_\alpha \}$ provides a 
lower bound of $c(\varrho)$, which can be tightened by numerical optimization.
However, all the examples we have considered so far suggest that there
is one matrix ${\cal T}^\alpha$ that 
gives the main contribution to the
rhs of eq. (\ref{lowb}).
Hence, the singular values of this matrix provide a purely algebraic
lower bound for $c$,
which often leads to satisfactory results even without further 
numerical refinement.
\begin{figure}
\epsfig{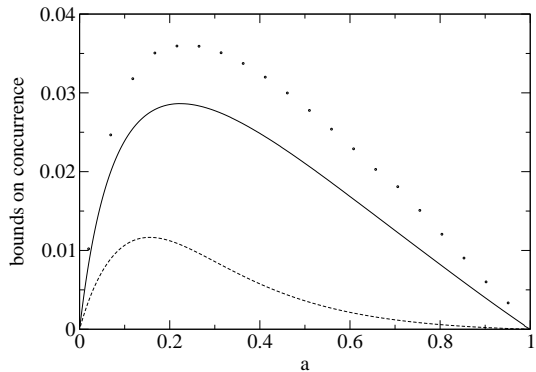}
\caption{Lower bound (full line) of  
the concurrence $c$ of the family of two spin-1 states $\varrho_a$
\protect\cite{hor97},  
together with its purely algebraic approximation 
$c_l^{(1)}=\lambda_1^{(1)}-\sum_{i>1}\lambda_i^{(1)}$ (dashed line), 
over the entire 
parameter range $a=0\ldots 1$.
Whilst $\varrho_a$ has positive partial transpose and is therefore 
not detected via the negative partial transpose criterion for entanglement,
already the algebraic approximation of our lower bound indentifies the state
as nonseparable, without need for numerical optimisation, for all $a$. Optimal
lower bound and optimal upper bound (dotted line, 
obtained by minimizing $\cal C$ in
(\ref{conc}) \cite{flo04}) confine the actual value of concurrence to an
interval with relative uncertainty 
of the order of approx. $10\%$.}
\label{hordat}
\end{figure}
As an example, consider 
the family 
of two spin-1 states 
\be
\varrho_a=\frac{1}{1+8a}\left[
{\renewcommand{\arraystretch}{0.8}\ba{ccccccccc}
a & 0 & 0 & 0 & a & 0 & 0 & 0 & a \\
0 & a & 0 & 0 & 0 & 0 & 0 & 0 & 0 \\
0 & 0 & a & 0 & 0 & 0 & 0 & 0 & 0 \\
0 & 0 & 0 & a & 0 & 0 & 0 & 0 & 0 \\
a & 0 & 0 & 0 & a & 0 & 0 & 0 & a \\
0 & 0 & 0 & 0 & 0 & a & 0 & 0 & 0 \\
0 & 0 & 0 & 0 & 0 & 0 & \beta & 0 & \gamma \\
0 & 0 & 0 & 0 & 0 & 0 & 0 & a & 0 \\
a & 0 & 0 & 0 & a & 0 & \gamma & 0 & \beta
\ea}\right],\ a\in[0,1]\ ,
\ee
with
$\beta=(1+a)/2$ and
$\gamma=\sqrt{1-a^2}/2$,
described in \cite{hor97}: The state 
$\varrho_a$ has positive partial transpose in the entire 
%parameter 
range of 
$a$, {\it i.e.}, 
the 
standard 
criterium \cite{per96} 
identifying a mixed state as nonseparable via negativity of its
partial transpose is inoperational here.
Notwithstanding,
the singular values of ${\cal T}^1$ associated with 
the largest eigenvalue of $B$ in eq.~(\ref{expandA}) above 
already provide a positive lower bound
$c_l^{(1)}=\lambda_1^{(1)}-\sum_{i>1}\lambda_i^{(1)}$ for $c(\varrho_a)$, as 
shown 
in
Fig.~\ref{hordat}:
$\varrho_a$ is detected as entangled without
any need 
for further numerical optimisation.
However, the figure also shows that taking 
into account all ${\cal T}^\alpha$, numerical optimisation significantly
raises  
the lower bound.
The remaining gap with respect to the upper bound
obtained from numerical minimization of $\cal C$ in (\ref{conc}) leaves a
relative uncertainty of the order of approx. $10\%$ on 
the actual
value of $c(\varrho)$, in the particularly pathological test case considered
here. 

Let us finally note that 
the original definition 
of concurrence \cite{wot98} is nicely embedded
in our formalism.
In the case of two-level systems one has $m=1$,
{\it i.e.}, there is only one non-vanishing matrix $T^1$.
Therefore, eq.~(\ref{capc}) simplifies to
\be
{\cal C}=\sum_{i=1}^N
\left|\left[VT^1V^T\right]_{ii}\right| \ ,
\ee
and the infimum can be 
derived analytically. Indeed, it is found 
that $T^1$ coincides with 
$\tau$ defined in the original work \cite{wot98}.

In conclusion, we have shown that a suitable representation of the 
concurrence of bipartite mixed quantum states in terms of the
eigensystem of a tensorial quantity allows for the derivation of a
lower bound of $c(\varrho)$, for arbitrary $\rho$.
Not only can this bound be tightened by an optimisation
under the comparatively simple constraint
$\sum_\alpha |z_\alpha|^2=1$, 
over a {\em complex} vector
space of dimension $n_1(n_1-1)n_2(n_2-1)/4$ 
-- at least a
factor $4n_1n_2$ smaller than dimensions of optimisation procedures
hitherto available.
It also can be reduced to a purely algebraic bound which appears to
provide good estimates, according to numerical data which
complement our analytical work.
Last but not least, our result 
can serve to derive lower bounds
on the entanglement of formation $E$ \cite{ben96} of a mixed quantum state,
thus quantifying the minimally required  
resources to
prepare $\varrho$:
Given any monotonously increasing, convex function $\cal E$ which satisfies 
${\cal E}(c(\psi))\le-\tr\varrho_r\log\varrho_r$, it follows that
$E(\varrho)\ge{\cal E}(c(\varrho))$, with the rhs bounded from below by our
bound (\ref{bound}).

We are indebted to Andr\'e Ricardo Ribeiro de Carvalho,
Rafa{\l} Demkowicz-Dobrza{\'n}ski and 
Karol {\.Z}yczkowski for fruitful discussions,
comments and remarks.
Financial support by VolkswagenStiftung and the Polish Ministery of Science
through the grant No. PBZ-MIN-008/P03/2003 is gratefully acknowledged.

\bibliography{referenzen}

\begin{thebibliography}{17}
\expandafter\ifx\csname natexlab\endcsname\relax\def\natexlab#1{#1}\fi
\expandafter\ifx\csname bibnamefont\endcsname\relax
  \def\bibnamefont#1{#1}\fi
\expandafter\ifx\csname bibfnamefont\endcsname\relax
  \def\bibfnamefont#1{#1}\fi
\expandafter\ifx\csname citenamefont\endcsname\relax
  \def\citenamefont#1{#1}\fi
\expandafter\ifx\csname url\endcsname\relax
  \def\url#1{\texttt{#1}}\fi
\expandafter\ifx\csname urlprefix\endcsname\relax\def\urlprefix{URL }\fi
\providecommand{\bibinfo}[2]{#2}
\providecommand{\eprint}[2][]{\url{#2}}

\bibitem[{\citenamefont{Nielsen and Chuang}(2000)}]{nielsen00}
\bibinfo{author}{\bibfnamefont{M.~A.} \bibnamefont{Nielsen}} \bibnamefont{and}
  \bibinfo{author}{\bibfnamefont{I.~L.} \bibnamefont{Chuang}},
  \emph{\bibinfo{title}{Quantum computation and quantum information}}
  (\bibinfo{publisher}{Cambridge University Press}, \bibinfo{year}{2000}).

\bibitem[{\citenamefont{Bru{\ss}}(2002)}]{bru02}
\bibinfo{author}{\bibfnamefont{D.}~\bibnamefont{Bru{\ss}}},
  \bibinfo{journal}{J. Math. Phys.} \textbf{\bibinfo{volume}{43}},
  \bibinfo{pages}{4237} (\bibinfo{year}{2002}).

\bibitem[{\citenamefont{Peres}(1996)}]{per96}
\bibinfo{author}{\bibfnamefont{A.}~\bibnamefont{Peres}},
  \bibinfo{journal}{Phys. Rev. Lett.} \textbf{\bibinfo{volume}{77}},
  \bibinfo{pages}{1413} (\bibinfo{year}{1996}).

\bibitem[{\citenamefont{Vidal and Werner}(2002)}]{vidal02}
\bibinfo{author}{\bibfnamefont{G.}~\bibnamefont{Vidal}} \bibnamefont{and}
  \bibinfo{author}{\bibfnamefont{R.~F.} \bibnamefont{Werner}},
  \bibinfo{journal}{Phys. Rev. A} \textbf{\bibinfo{volume}{65}},
  \bibinfo{pages}{032314} (\bibinfo{year}{2002}).

\bibitem[{\citenamefont{Terhal}(2000)}]{ter00b}
\bibinfo{author}{\bibfnamefont{B.}~\bibnamefont{Terhal}},
  \bibinfo{journal}{Phys. Lett. A} \textbf{\bibinfo{volume}{271}},
  \bibinfo{pages}{319} (\bibinfo{year}{2000}).

\bibitem[{\citenamefont{Doherty et~al.}(2002)\citenamefont{Doherty, Parrilo,
  and Spedalieri}}]{doh02}
\bibinfo{author}{\bibfnamefont{A.~C.} \bibnamefont{Doherty}},
  \bibinfo{author}{\bibfnamefont{P.~A.} \bibnamefont{Parrilo}},
  \bibnamefont{and} \bibinfo{author}{\bibfnamefont{F.~M.}
  \bibnamefont{Spedalieri}}, \bibinfo{journal}{Phys. Rev. Lett.}
  \textbf{\bibinfo{volume}{88}}, \bibinfo{pages}{187904}
  (\bibinfo{year}{2002}).

\bibitem[{\citenamefont{Hughston et~al.}(1993)\citenamefont{Hughston, Josza,
  and Wootters}}]{wot93}
\bibinfo{author}{\bibfnamefont{L.~P.} \bibnamefont{Hughston}},
  \bibinfo{author}{\bibfnamefont{R.}~\bibnamefont{Josza}}, \bibnamefont{and}
  \bibinfo{author}{\bibfnamefont{W.~K.} \bibnamefont{Wootters}},
  \bibinfo{journal}{Phys. Lett. A} \textbf{\bibinfo{volume}{183}},
  \bibinfo{pages}{14} (\bibinfo{year}{1993}).

\bibitem[{\citenamefont{Bennet et~al.}(1996)\citenamefont{Bennet, DiVincenzo,
  Smolin, and Wootters}}]{ben96}
\bibinfo{author}{\bibfnamefont{C.~H.} \bibnamefont{Bennet}},
  \bibinfo{author}{\bibfnamefont{D.~P.} \bibnamefont{DiVincenzo}},
  \bibinfo{author}{\bibfnamefont{J.}~\bibnamefont{Smolin}}, \bibnamefont{and}
  \bibinfo{author}{\bibfnamefont{W.~K.} \bibnamefont{Wootters}},
  \bibinfo{journal}{Phys. Rev. A} \textbf{\bibinfo{volume}{54}},
  \bibinfo{pages}{3824} (\bibinfo{year}{1996}).

\bibitem[{\citenamefont{Wootters}(1998)}]{wot98}
\bibinfo{author}{\bibfnamefont{W.~K.} \bibnamefont{Wootters}},
  \bibinfo{journal}{Phys. Rev. Lett.} \textbf{\bibinfo{volume}{80}},
  \bibinfo{pages}{2245} (\bibinfo{year}{1998}).

\bibitem[{\citenamefont{Uhlmann}(2000)}]{uhl00}
\bibinfo{author}{\bibfnamefont{A.}~\bibnamefont{Uhlmann}},
  \bibinfo{journal}{Phys. Rev. A} \textbf{\bibinfo{volume}{62}},
  \bibinfo{pages}{032307} (\bibinfo{year}{2000}).

\bibitem[{\citenamefont{Rungta et~al.}(2001)\citenamefont{Rungta, Buzek, Caves,
  Hillery, Milburn, and Wootters}}]{run01}
\bibinfo{author}{\bibfnamefont{P.}~\bibnamefont{Rungta}},
  \bibinfo{author}{\bibfnamefont{V.}~\bibnamefont{Buzek}},
  \bibinfo{author}{\bibfnamefont{C.~M.} \bibnamefont{Caves}},
  \bibinfo{author}{\bibfnamefont{M.}~\bibnamefont{Hillery}},
  \bibinfo{author}{\bibfnamefont{G.~J.} \bibnamefont{Milburn}},
  \bibnamefont{and} \bibinfo{author}{\bibfnamefont{W.~K.}
  \bibnamefont{Wootters}}, \bibinfo{journal}{Phys. Rev. A}
  \textbf{\bibinfo{volume}{64}}, \bibinfo{pages}{042315}
  (\bibinfo{year}{2001}).

\bibitem[{\citenamefont{Uhlmann}(1998)}]{uhl98}
\bibinfo{author}{\bibfnamefont{A.}~\bibnamefont{Uhlmann}},
  \bibinfo{journal}{Open Sys. Info. Dyn.} \textbf{\bibinfo{volume}{5(3)}},
  \bibinfo{pages}{209} (\bibinfo{year}{1998}).

\bibitem[{\citenamefont{Schr{\"o}dinger}(1936)}]{schroed}
\bibinfo{author}{\bibfnamefont{E.}~\bibnamefont{Schr{\"o}dinger}},
  \bibinfo{journal}{Proc. Cambridge Philos. Soc} \textbf{\bibinfo{volume}{32}},
  \bibinfo{pages}{446} (\bibinfo{year}{1936}).

\bibitem[{\citenamefont{Badzi\c{a}g et~al.}(2002)\citenamefont{Badzi\c{a}g,
  Deuar, Horodecki, Horodecki, and Horodecki}}]{bad02}
\bibinfo{author}{\bibfnamefont{P.}~\bibnamefont{Badzi\c{a}g}},
  \bibinfo{author}{\bibfnamefont{P.}~\bibnamefont{Deuar}},
  \bibinfo{author}{\bibfnamefont{M.}~\bibnamefont{Horodecki}},
  \bibinfo{author}{\bibfnamefont{P.}~\bibnamefont{Horodecki}},
  \bibnamefont{and}
  \bibinfo{author}{\bibfnamefont{R.}~\bibnamefont{Horodecki}},
  \bibinfo{journal}{J. Mod. Opt} \textbf{\bibinfo{volume}{49}},
  \bibinfo{pages}{1289} (\bibinfo{year}{2002}).

\bibitem[{\citenamefont{Audenaert et~al.}(2001)\citenamefont{Audenaert,
  Verstraete, and Moor}}]{aud01}
\bibinfo{author}{\bibfnamefont{K.}~\bibnamefont{Audenaert}},
  \bibinfo{author}{\bibfnamefont{F.}~\bibnamefont{Verstraete}},
  \bibnamefont{and} \bibinfo{author}{\bibfnamefont{B.~D.} \bibnamefont{Moor}},
  \bibinfo{journal}{Phys. Rev. A} \textbf{\bibinfo{volume}{64}},
  \bibinfo{pages}{052304} (\bibinfo{year}{2001}).

\bibitem[{\citenamefont{Horodecki}(1997)}]{hor97}
\bibinfo{author}{\bibfnamefont{P.}~\bibnamefont{Horodecki}},
  \bibinfo{journal}{Phys. Lett. A} \textbf{\bibinfo{volume}{232}},
  \bibinfo{pages}{333} (\bibinfo{year}{1997}).

\bibitem[{\citenamefont{Mintert}(2004)}]{flo04}
\bibinfo{author}{\bibfnamefont{F.}~\bibnamefont{Mintert}}, Ph.D. thesis,
  \bibinfo{school}{Ludwig-Maximilians-Universit\"at M\"unchen (submitted)}
  (\bibinfo{year}{2004}).

\end{thebibliography}

\end{document}